\documentclass[a4paper,12pt,reqno]{article}
\usepackage{amsmath,amsfonts,amssymb,amsthm,dsfont}

\usepackage{hyperref}
\allowdisplaybreaks
\numberwithin{equation}{section}

\begin{document}

\begin{flushright}
ITP--UH--02/12
\end{flushright}

\begin{center}
 {\large\bfseries Aspects of supersymmetric BRST cohomology}
 \\[5mm]
 Friedemann Brandt \\[2mm]
 \textit{Institut f\"ur Theoretische Physik, Leibniz Universit\"at Hannover, Appelstra\ss e 2, D-30167 Hannover, Germany}
\end{center}

\begin{abstract}
The application and extension of well-known BRST cohomological methods to supersymmetric field theories are discussed. The focus is on the emergence and particular features of supersymmetry algebra cohomology in this context. In particular it is discussed and demonstrated that supersymmetry algebra cohomology emerges within the cohomological analysis of standard supersymmetric field theories whether or not the commutator algebra of the symmetry transformations closes off-shell.
\end{abstract}

\section{Introduction}\label{sec1}

The BRST formalism was originally developed in gauge theories of the Yang-Mills type \cite{Becchi:1974md,Zinn-Justin:1974mc,Becchi:1975nq,Tyutin:1975qk} and later extended to gauge theories with gauge transformations whose commutator algebra closes only on-shell (so-called ``open algebras")  \cite{Kallosh:1978de,deWit:1978cd}. The outcome of these developments was an elegant and universal formalism \cite{Batalin:1981jr} usable to construct nilpotent BRST-type transformations for a generic Lagrangean gauge theory. This formalism is nowadays often called BRST-antifield, field-antifield or BV formalism. It was further extended so as to include global symmetries \cite{Brandt:1997cz} (``extended antifield formalism"). We shall call this general concept simply BRST formalism, the respective nilpotent transformations of the fields and antifields BRST transformations and the operation implementing these transformations BRST differential denoted by $s$.
 
The nilpotency of the BRST differential ($s^2=0$) allows one to define the local BRST cohomology as the cohomology of $s$ on local functions or differential forms constructed of fields and antifields (or the relative cohomology of $s$ modulo total derivatives or $d$-exact local differential forms, see below). This cohomology has useful applications both in quantum field theory and in classical field theory. In quantum field theory it determines candidate counterterms and anomalies \cite{Weinberg:1996kr}. In classical field theory it determines consistent deformations of invariant actions and their symmetries \cite{Barnich:1993vg,Brandt:1998gj}, and local conservation laws \cite{Barnich:1994db}.

Maximilian Kreuzer contributed substantially to the computation of the antifield independent local BRST cohomology in Yang-Mills and gravitational theories \cite{Brandt:1989rd,Brandt:1989gy,Brandt:1989gv,Brandt:1989et} which provided methods that meanwhile have become standard tools to investigate the local BRST cohomology.
The purpose of the present contribution is to explain how these methods can be applied and extended to ``supersymmetric BRST cohomology" which is an abbreviation for the local BRST cohomology in supersymmetric field theories when $s$ contains the supersymmetry transformations. The focus is on the emergence and particular features of what we call supersymmetry algebra cohomology (SAC) in this context.

Sections \ref{sec2} and \ref{sec3} briefly review the definition of SAC and a method to compute it systematically, respectively. In section \ref{sec4} the emergence of SAC within the BRST cohomological analysis is discussed. Section \ref{sec5} comments on the so-called descent equations in supersymmetric BRST cohomology. Section \ref{sec6} demonstrates typical aspects of SAC by means of an example which is simple but yet suitable to illustrate these aspects. Section \ref{sec7} contains a few concluding remarks.

\section{Supersymmetry algebra cohomology (SAC)}\label{sec2}

SAC \cite{Brandt:2009xv} is related to standard
supersymmetry algebras (SUSY algebras) of bosonic translational generators $P_a$ ($a=1,\dots,D$) and fermionic supersymmetry generators $Q^i_{\underline{\alpha}}$ (where ${\underline{\alpha}}$ is a spinor index and $i=1,\dots,N$) in $D$-dimensional spacetime of the form
 \begin{align}
  [\,P_a\, ,\, P_b\, ]=0,\quad [\, P_a\, ,\, Q^i_{\underline{\alpha}}\, ]=0,\quad \{Q^i_{\underline{\alpha}}\, ,\, Q^j_{\underline{\beta}}\}=M^{ij}\,(\Gamma^a C^{-1})_{{\underline{\alpha}}{\underline{\beta}}}P_a
  \label{alg}
 \end{align}
where $[\,A\, ,\, B\, ]=AB-BA$ denotes the commutator of two generators $A$ and $B$, $\{A\, ,\, B\}=AB+BA$ denotes the anticommutator of two generators $A$ and $B$, $\Gamma^a$ are gamma-matrices in $D$ dimensions and $C$ is a related charge conjugation matrix, and $M^{ij}$ are the entries of a (generally complex) $N\times N$ matrix. 
The supersymmetry algebra \eqref{alg} is represented on variables $\tilde T$ constructed of fields and antifields which we shall discuss later on.

The SAC related to a supersymmetry algebra \eqref{alg} is defined by means of a BRST-type differential $s_\mathrm{\,susy}$ constructed of the generators $P_a$, $Q^i_{\underline{\alpha}}$ of the supersymmetry algebra and corresponding fermionic ghost variables $c^a$ (``translation ghosts") and bosonic ghost variables $\xi^{\underline{\alpha}}_i$ (``supersymmetry ghosts") according to
 \begin{align}
  s_\mathrm{\,susy}=c^a P_a+\xi^{\underline{\alpha}}_i \,Q_{\underline{\alpha}}^i-\tfrac{1}{2} M^{ij}(\Gamma^a C^{-1})_{{\underline{\alpha}}{\underline{\beta}}}\,\xi^{\underline{\alpha}}_i\xi^{\underline{\beta}}_j\,\frac{\partial}{\partial c^a}
	\label{brs} 
 \end{align}
where the generators $P_a$ and $Q_{\underline{\alpha}}^i$ act nontrivially only on the variables $\tilde T$. On functions $\omega(c,\xi,\tilde T)$ of the variables $c,\xi,\tilde T$, the differential $s_\mathrm{\,susy}$ is defined as an antiderivation and, therefore, by construction squares to zero (($s_\mathrm{\,susy})^2\omega=0$). This allows one to define the SAC as the cohomology $H(s_\mathrm{\,susy})$ in a space $\Omega$ of functions $\omega(c,\xi,\tilde T)$ of the variables $c,\xi,\tilde T$ where the dependence on the ghosts $c,\xi$ is always polynomial (the dependence on variables $\tilde T$ or a subset thereof may be nonpolynomial, depending on the context),
 \begin{align}
  H(s_\mathrm{\,susy})= \frac{\mathrm{kernel\ of\ }s_\mathrm{\,susy}\ \mathrm{in}\ \Omega}{\mathrm{image\ of\ }s_\mathrm{\,susy}\ \mathrm{in}\ \Omega}\ .
  \label{brscoh}
 \end{align}
The representatives of $H(s_\mathrm{\,susy})$ are thus elements $\omega\in\Omega$ which fulfill  $s_\mathrm{\,susy}\,\omega=0$, and two elements $\omega,\omega'\in\Omega$ are considered equivalent in $H(s_\mathrm{\,susy})$ if $\omega'=\omega+s_\mathrm{\,susy}\,\eta$ for some $\eta\in\Omega$.
 
\section{SUSY ladder equations}\label{sec3}

To compute a SAC systematically it is useful to decompose the cocycle condition $s_\mathrm{\,susy}\,\omega=0$ with respect to the degree in the translation ghosts, analogously to a strategy that was used in standard gravitational theories \cite{Brandt:1989et,Barnich:1995ap}. We call the degree in the translation ghosts $c$-degree\ and denote the operator which counts the translation ghosts by $N_c=c^a\frac{\partial}{\partial c^a}$. $s_\mathrm{\,susy}$ decomposes into three parts $d_c$, $d_\xi$, $s_\mathrm{gh}$ which have $c$-degree\ $+1$, $0$ and $-1$, respectively, i.e. $d_c$ increments the $c$-degree\ by one, $d_\xi$ does not change the $c$-degree\ and $s_\mathrm{gh}$ decrements the $c$-degree\ by one:
\begin{align}
&s_\mathrm{\,susy}=d_c+d_\xi+s_\mathrm{gh}\ ,\notag\\
&d_c=c^a P_a\,,\ d_\xi=\xi^{\underline{\alpha}}_i Q_{\underline{\alpha}}^i\,,\ 
s_\mathrm{gh}=-\tfrac{1}{2}\, M^{ij}\,(\Gamma^a C^{-1})_{{\underline{\alpha}}{\underline{\beta}}}\,\xi^{\underline{\alpha}}_i\xi^{\underline{\beta}}_j\,\frac{\partial}{\partial c^a}\ .
\label{gen8} 
\end{align}
An element $\omega\in\Omega$ decomposes into parts $\omega^p\in\Omega$ with various $c$-degree s $p$,
\begin{align}
\omega=\sum_{p=m}^{M}\omega^{p},\quad N_c\,\omega^{p}=p\,\omega^{p}\ .
\end{align}
The cocycle condition $s_\mathrm{\,susy}\,\omega=0$ decomposes into a tower of equations which we call
SUSY ladder equations:
\begin{align}
s_\mathrm{\,susy}\,\omega=0\ \Leftrightarrow\ 
\left\{\begin{array}{l}
0=s_\mathrm{gh}\,\omega^{m}\\
0=d_\xi\,\omega^{m}+s_\mathrm{gh}\,\omega^{m+1}\\
0=d_c\,\omega^{p}+d_\xi\,\omega^{p+1}+s_\mathrm{gh}\,\omega^{p+2}\ \mathrm{for}\ m\leq p\leq M-2\\
0=d_c\,\omega^{M-1}+d_\xi\,\omega^{M}\\
0=d_c\,\omega^{M}
\end{array}
\right.
\label{ladder}
\end{align}
The SUSY ladder equations provide a systematic method to analyse the SAC by relating $H(s_\mathrm{\,susy})$ to $H(s_\mathrm{gh})$, i.e. to the cohomology of the part $s_\mathrm{gh}$ of $s_\mathrm{\,susy}$. Indeed, the part $\omega^{m}$ with lowest $c$-degree\ contained in a solution of $s_\mathrm{\,susy}\,\omega=0$ solves $s_\mathrm{gh}\,\omega^m=0$ because $s_\mathrm{gh}$ is the only part of $s_\mathrm{\,susy}$ which decrements the $c$-degree. Hence, $\omega^m$ is a cocycle in $H(s_\mathrm{gh})$. By means of the SUSY ladder equations one can thus relate $H(s_\mathrm{\,susy})$ to $H(s_\mathrm{gh})$ using spectral sequence techniques. $s_\mathrm{gh}$ only involves the ghost variables and the structure constants $M^{ij}\,(\Gamma^a C^{-1})_{{\underline{\alpha}}{\underline{\beta}}}$ of the supersymmetry algebra (\ref{alg}). In particular, $H(s_\mathrm{gh})$ does not depend on the way the supersymmetry algebra (\ref{alg}) is represented on variables $\tilde T$ and is thus the ``universal" part of $H(s_\mathrm{\,susy})$. It should be noted, however, that $H(s_\mathrm{gh})$ depends on the dimension $D$ and on the number $N$ of sets of supersymmetries \cite{Brandt:2010fa,Brandt:2010tz,Movshev:2010mf,Movshev:2011pr}.

\section{Emergence of SAC in supersymmetric BRST cohomology}\label{sec4}

In practically all relevant supersymmetric field theories the commutators of the symmetry transformations do not directly provide a SUSY algebra \eqref{alg}. In particular the commutators of supersymmetry transformations normally contain symmetry transformations different from translations (such as Yang-Mills gauge transformations) and/or trivial symmetry transformations which vanish only on-shell and lead to so-called ``open algebras" (i.e. commutator algebras which close only on-shell). Furthermore, in supergravity theories the ghosts $c$ and $\xi$ are ghost fields and the BRST transformations contain also derivatives of these ghost fields. Nevertheless the simple SAC as defined above emerges within the local BRST cohomology of standard supersymmetric field theories, whether or not the commutator algebra of the symmetry transformations closes off-shell or commutators of supersymmetry transformations contain symmetry transformations different from translations or the BRST transformations involve derivatives of ghost fields $c$ or $\xi$. This section is to indicate why and how this happens, without going into details.

The emergence of the SAC in the local BRST cohomology is related to a method which is sometimes called ``elimination of trivial pairs". This method ``eliminates" pairs of variables $u,v$ forming ``BRST doublets" ($v=s\,u$) from the cohomological analysis. It was introduced already within the antifield independent cohomological analyses of standard Yang-Mills and gravitational theories \cite{Brandt:1989rd,Brandt:1989gy,Brandt:1989et} and used, among others, to eliminate the derivatives of the ghosts from the cohomology. In a cohomological analysis which includes the antifields this elimination of trivial pairs involving derivatives of the ghosts reduces the cohomological problem to a problem involving only ghost variables corresponding to the undifferentiated ghosts, tensors $T$ of standard type and ``antitensors" $T^\star$ corresponding to the antifields and their derivatives \cite{Barnich:1995ap}.

In standard field theories 
this method can be extended so as to eliminate also the antifields and their derivatives from the cohomological analysis along with tensors which vanish on-shell \cite{Brandt:1996mh,Brandt:2001tg}. This further reduces the cohomological problem to a problem involving only specific ghost variables and ``generalized tensors" $\tilde T=T+\dots$, with the ellipsis indicating possible contributions depending on antifields. In a standard supersymmetric field theory a SAC arises within this reduced cohomological problem by linearizing the BRST-transformations in the generalized tensors $\tilde T$. The resultant cohomological problem involves a SUSY algebra \eqref{alg} which is represented on the generalized tensors $\tilde T$, even when the symmetry transformations form an open algebra (because tensors which vanish on-shell are eliminated from the cohomological analysis along with the antifields). Accordingly, in this approach antifields enter the representatives of the local BRST cohomology only via the antifield dependent parts of the generalized tensors $\tilde T$. Furthermore the approach implicates that the translational generators $P_a$ of the SUSY algebra \eqref{alg} are represented on the generalized tensors $\tilde T$ in an unusual way which corresponds to a representation of derivatives {\em on-shell}. We shall not further discuss these matters here but refer to section \ref{sec6} for an explicit example demonstrating this approach.

\section{Descent equations in supersymmetric BRST cohomology}\label{sec5}

We shall now discuss a feature of supersymmetric BRST cohomology which is completely analogous to a feature of standard gravitational theories \cite{Brandt:1989et,Barnich:1995ap} and is therefore only briefly sketched.
Many applications of local BRST cohomology are not directly related to the cohomology $H(s)$ of $s$ but to the so-called relative cohomology $H(s|d)$ of $s$ and the exterior derivative $d=dx^m\partial_m$ on local differential forms of the fields, antifields and their derivatives. The latter cohomology gives rise to so-called descent equations for $s$ and $d$ relating it to the cohomology $H(s+d)$ of $s+d$ on sums $\tilde \omega$ of local differential forms $\omega_p$ with different form-degrees $p$\,:
\begin{align}
\left.
\begin{array}{rcl}
s\,\omega_D+d\,\omega_{D-1}&=&0\\
s\,\omega_{D-1}+d\,\omega_{D-2}&=&0\\
&\vdots&\\
s\,\omega_k&=&0
\end{array}
\right\}
\Leftrightarrow\ 
(s+d)\,\tilde\omega=0,\ \tilde\omega=\sum_{p=k}^D\omega_p\ .
\label{descent}
\end{align}
These descent equations relate $H(s|d)$ and $H(s+d)$ to $H(s)$. In general these relations can be quite subtle as in the case of Yang-Mills theories. However, they are direct in the local BRST cohomology of standard gravitational or supersymmetric field theories. The reason is that in these cases $s$ contains a part $c^m\partial_m$ where $c^m$ are diffeomorphism or translation ghosts and $\partial_m$ differentiates the fields and antifields, and this is the only occurrence of the undifferentiated ghosts $c^m$ in the BRST transformations. Therefore, when acting on fields and antifields, $s+d$ arises from $s$ simply by substituting $c^m+dx^m$ for $c^m$. As a consequence, in these theories practically all relevant information on $H(s|d)$ and $H(s+d)$ is already contained in $H(s)$. Furthermore the representatives of $H(s)$ directly provide representatives of $H(s+d)$ by substituting $c^m+dx^m$ for $c^m$ everywhere. 

\section{Example in two dimensions}\label{sec6}

\subsection{Model}\label{sec6.1}

We shall now illustrate some aspects of SAC discussed above by means of a simple globally supersymmetric field theory (``model") in flat two-dimensional spacetime with Minkowski metric $\eta_{ab}=\mathrm{diag}(-1,+1)$. In order to make formulae explicit we use the gamma-matrices $\Gamma_1\equiv\mathrm{i}\,\sigma_1$, $\Gamma_2\equiv\sigma_2$ and the charge conjugation matrix $C\equiv\sigma_2$ where $\sigma_1$ and $\sigma_2$ denote the first and second Pauli matrix, respectively. For real Majorana supersymmetry generators $(Q_{\underline{1}},Q_{\underline{2}})\equiv (Q_+,Q_-)$ (two real Majorana-Weyl supersymmetries $(Q_+,0)$ and $(0,Q_-)$ of opposite chirality) the SUSY algebra \eqref{alg} reads explicitly (with $M^{ij}=-\mathrm{i}\,\delta^{ij}$):
\begin{align}
  &[\,P_1\, ,\, P_2\,]=0,\ [\,P_1\, ,\, Q_\pm\,]=0,\ [\, P_2\, ,\, Q_\pm\,]=0,\notag\\ &
  (Q_+)^2=-\tfrac{\mathrm{i}}{2}\,(P_1+P_2),\ (Q_-)^2=\tfrac{\mathrm{i}}{2}\,(P_1-P_2),\ \{Q_+\, ,\, Q_-\}=0.
  \label{2Dalg}
\end{align}
The model to be discussed is a free field theory 
for a real bosonic field $\varphi$ and a fermionic field $\psi$ with real spinor components $(\psi_{\underline{1}},\psi_{\underline{2}})\equiv(\psi_+,\psi_-)$. The Lagrangian of the model reads 
\begin{align}
L&=-\tfrac{1}{2}\,\eta^{ab}\partial_a\varphi\,\partial_b\varphi-\mathrm{i}\,\psi^{\underline{\alpha}}(\Gamma^a C^{-1})_{{\underline{\alpha}}{\underline{\beta}}}\,\partial_a\psi^{\underline{\beta}}\notag\\
&=\tfrac{1}{2}\,(\partial_1\varphi)^2-\tfrac{1}{2}\,(\partial_2\varphi)^2+\mathrm{i}\,\psi_-(\partial_1+\partial_2)\psi_--\mathrm{i}\,\psi_+(\partial_1-\partial_2)\psi_+\, .
\label{L}
\end{align}
This Lagrangian is invariant up to total derivatives under infinitesimal symmetry transformations $\delta_a$ and $\delta_{\underline{\alpha}}$ given by
\begin{align}
\delta_a\varphi=\partial_a\varphi\,,\ \delta_a\psi_{\underline{\alpha}}=\partial_a\psi_{\underline{\alpha}}\,,\ 
\delta_{\underline{\alpha}}\varphi=\psi_{\underline{\alpha}}\,,\ \delta_{\underline{\alpha}}\psi_{\underline{\beta}}=-\tfrac{\mathrm{i}}{2}\,(\Gamma^a C^{-1})_{{\underline{\alpha}}{\underline{\beta}}}\,\partial_a\varphi\, .
\label{symm1}
\end{align}
The transformations $(\delta_{\underline{1}},\delta_{\underline{2}})\equiv (\delta_+,\delta_-)$ are supersymmetry transformations which read explicitly
\begin{align}
\delta_\pm\varphi=\psi_\pm\,,\ \delta_\pm\psi_\pm=\mp\tfrac{\mathrm{i}}{2} (\partial_1\pm\partial_2)\varphi\,,\
\delta_\pm\psi_\mp=0.
\label{symm2}
\end{align}
It can be readily verified that the graded commutator algebra of the transformations $\delta_a$ and $\delta_{\underline{\alpha}}$ provides an on-shell representation of the SUSY algebra \eqref{2Dalg} with $P_a$ and $Q_{\underline{\alpha}}$ represented by $\delta_a$ and $\delta_{\underline{\alpha}}$\,, respectively. For instance, one has $(\delta_+)^2\psi_-=0$ which is equal to $-\tfrac{\mathrm{i}}{2}\,(\partial_1+\partial_2)\psi_-$ on-shell because $(\partial_1+\partial_2)\psi_-$ vanishes on-shell (by the equation of motion for $\psi_-$).

\subsection{BRST transformations for the model}\label{sec6.2}

The BRST transformations corresponding to the symmetry transformations $\delta_a$ and $\delta_{\underline{\alpha}}$ for the model read
\begin{align}
s\,\varphi&=c^a\partial_a\varphi+\xi^+\psi_++\xi^-\psi_-\, ,\notag\\
s\,\psi_+&=c^a\partial_a\psi_+-\tfrac{\mathrm{i}}{2}\,\xi^+\partial_+\varphi+\tfrac{1}{4}\,\xi^-\xi^-\psi^{\star +}-\tfrac{1}{4}\,\xi^+\xi^-\psi^{\star -}\, ,\notag\\
s\,\psi_-&=c^a\partial_a\psi_-+\tfrac{\mathrm{i}}{2}\,\xi^-\partial_-\varphi+\tfrac{1}{4}\,\xi^+\xi^+\psi^{\star -}-\tfrac{1}{4}\,\xi^+\xi^-\psi^{\star +}\, ,\notag\\
s\,\varphi^\star&=-\partial_+\partial_-\varphi+c^a\partial_a\varphi^\star-\tfrac{\mathrm{i}}{2}\,\xi^+\partial_+\psi^{\star +}+\tfrac{\mathrm{i}}{2}\,\xi^-\partial_-\psi^{\star -}\, ,\notag\\
s\,\psi^{\star +}&=2\mathrm{i}\,\partial_-\psi_++c^a\partial_a\psi^{\star +}+\xi^+\varphi^\star\, ,\notag\\
s\,\psi^{\star -}&=-2\mathrm{i}\,\partial_+\psi_-+c^a\partial_a\psi^{\star -}+\xi^-\varphi^\star\, ,\notag\\
s\,c^+&=\mathrm{i}\,\xi^+\xi^+\, ,\ 
s\,c^-=-\mathrm{i}\,\xi^-\xi^-\, ,\ 
s\, \xi^+=0\, ,\  s\, \xi^-=0 
\label{strafos}
\end{align}
where $\varphi^\star$, $\psi^{\star +}$, $\psi^{\star -}$ are the antifields corresponding to $\varphi$, $\psi_+$, $\psi_-$, respectively, the $c$ and $\xi$ are constant ghosts, and
\begin{equation}
\partial_\pm=\partial_1\pm\partial_2\,,\ c^\pm=c^1\pm c^2,\ c^a\partial_a=c^1\partial_1+c^2\partial_2=\tfrac{1}{2}(c^+\partial_++c^-\partial_-).
\label{change}
\end{equation}

\subsection{Trivial pairs and generalized tensors}\label{sec6.3}

We shall now explain how the ``elimination of trivial pairs" outlined in section \ref{sec4} works in the model under consideration. In the present case the antifields $\varphi^\star$, $\psi^{\star +}$, $\psi^{\star -}$ and their derivatives are variables $u^\ell$ of BRST-doublets $(u^\ell,v^\ell)$ with $v^\ell=s\,u^\ell$:
\begin{align}
&\{u^\ell\}=\{\partial_+^m\partial_-^n\varphi^\star,\ \partial_+^m\partial_-^n\psi^{\star +},\ \partial_+^m\partial_-^n\psi^{\star -}\,|\, m,n=0,1,2,\dots\}\, ,\notag\\
&s\,\partial_+^m\partial_-^n\varphi^\star=-\partial_+^{m+1}\partial_-^{n+1}\varphi+\dots\ ,\notag\\
&s\,\partial_+^m\partial_-^n\psi^{\star +}=2\mathrm{i}\,\partial_+^{m}\partial_-^{n+1}\psi_++\dots\ ,\notag\\
&s\,\partial_+^m\partial_-^n\psi^{\star -}=-2\mathrm{i}\,\partial_+^{m+1}\partial_-^{n}\psi_-+\dots
\label{uv}
\end{align}
with ellipsis indicating antifield dependent terms, and $\partial_\pm^m=(\partial_\pm)^m=\partial_\pm\dots\partial_\pm$. The variables $v^\ell$ substitute for all those derivatives of $\varphi$ and $\psi_{\underline{\alpha}}$ which vanish on-shell, i.e. which are set to zero by the equations of motion deriving from \eqref{L}. The undifferentiated fields $\varphi$ and $\psi_{\underline{\alpha}}$ and their remaining derivatives give rise to ``generalized tensors" $\tilde T_A$ which are constructed such that $s\,\tilde T_A=r_A(c,\xi,\tilde T)$ \cite{Brandt:2001tg}:
\begin{align}
&\{\tilde T_A\}=\{\varphi_{(0,0)},\, \varphi_{(m+1,0)},\, \varphi_{(0,m+1)},\, \psi_{+(m,0)},\, \psi_{-(0,m)}\,|\, m=0,1,2,\dots\}\, ,\notag\\
&\varphi_{(0,0)}=\varphi\, ,\notag\\
&\varphi_{(m+1,0)}=\partial_+^{m}(\partial_+\varphi-\tfrac{\mathrm{i}}{2}\,\xi^-\psi^{\star -}+\tfrac{1}{2}\, c ^-\varphi^\star)\, ,\notag\\
&\varphi_{(0,m+1)}=\partial_-^{m}(\partial_-\varphi+\tfrac{\mathrm{i}}{2}\,\xi^+\psi^{\star +}-\tfrac{1}{2}\, c ^+\varphi^\star)\, ,\notag\\
&\psi_{+(m,0)}=\partial_+^{m}(\psi_+-\tfrac{\mathrm{i}}{4}\,c^-\psi^{\star +})\, ,\notag\\
&\psi_{-(0,m)}=\partial_-^{m}(\psi_-+\tfrac{\mathrm{i}}{4}\,c^+\psi^{\star -})\, .
\label{T}
\end{align}

\subsection{BRST transformations of the {\boldmath $\tilde T$} and SUSY algebra}\label{sec6.4}
 
In the present case
the BRST transformations of the $\tilde T$ are linear in the $\tilde T$ and in the ghosts. 
Hence, they can be written as
\begin{equation}
s\,\tilde T_A=(\tfrac{1}{2}\,c^+P_++\tfrac{1}{2}\,c^-P_-+\xi^+Q_++\xi^-Q_-)\,\tilde T_A
\label{sT}
\end{equation}
with linear transformations $P_\pm$ and $Q_\pm$ of the $\tilde T$. Owing to $s^2=0$ these transformations $P_\pm,Q_\pm$ have a graded commutator algebra whose structure constants can be read off from the BRST transformations of the ghosts. Of course, this graded commutator algebra is precisely the SUSY algebra \eqref{2Dalg} written in terms of $P_\pm=P_1\pm P_2$ as we have anticipated already by the notation,
\begin{align*}
&[\, P_+\, ,\, P_-\,]=[\, P_+\, ,\, Q_\pm\, ]=[\, P_-\, ,\, Q_\pm\, ]=0,\\ 
&(Q_+)^2=-\tfrac{\mathrm{i}}{2}\,P_+\,,\quad (Q_-)^2=\tfrac{\mathrm{i}}{2}\,P_-\,,\quad \{Q_+\, ,\, Q_-\}=0.
\end{align*}
The BRST transformations of the $\tilde T$ thus provide a representation of the SUSY algebra \eqref{2Dalg} with $P_\pm\,\tilde T_A$ and $Q_\pm\,\tilde T_A$ given by the coefficients of $c^\pm$ and $\xi^\pm$ in $s\,\tilde T_A$. Explicitly one obtains
\begin{equation}
\begin{array}{|c||c|c|c|c|c|}
 \hline\rule{0em}{2.5ex}
 \tilde T_A & \varphi_{(0,0)} & \varphi_{(m+1,0)} & \varphi_{(0,m+1)} & \psi_{+(m,0)} & \psi_{-(0,m)}\\
 \hline\rule{0em}{2.5ex}
 P_+\tilde T_A & \varphi_{(1,0)} & \varphi_{(m+2,0)} & 0 & \psi_{+(m+1,0)} & 0\\
 \hline\rule{0em}{2.5ex}
 P_-\tilde T_A & \varphi_{(0,1)} & 0 & \varphi_{(0,m+2)} & 0 &  \psi_{-(0,m+1)}\\
 \hline\rule{0em}{2.5ex}
 Q_+\tilde T_A & \psi_{+(0,0)} & \psi_{+(m+1,0)} & 0 & -\tfrac{\mathrm{i}}{2}\,\varphi_{(m+1,0)} & 0\\
 \hline\rule{0em}{2.5ex}
 Q_-\tilde T_A & \psi_{-(0,0)} & 0 & \psi_{-(0,m+1)} & 0 & \tfrac{\mathrm{i}}{2}\,\varphi_{(0,m+1)}\\
 \hline
 \end{array}
 \label{rep}
\end{equation}
As we have pointed out already at the end of section \ref{sec4}, the representation of the translational generators $P$ on the $\tilde T$ corresponds to a representation of derivatives on-shell. For instance, the generalized tensor $\varphi_{(1,0)}$ is mapped to zero by $P_-$ according to \eqref{rep}. This corresponds to the fact that
$\partial_-\partial_+\varphi$ vanishes on-shell because the equation of motion for $\varphi$ deriving from \eqref{L} sets $\partial_-\partial_+\varphi=(\partial_1)^2\varphi-(\partial_2)^2\varphi$ to zero.

\subsection{Computation of the supersymmetric BRST cohomology}\label{sec6.5}

In the following we sketch the computation of the supersymmetric BRST cohomology $H(s)$ for the model under consideration. As discussed in section \ref{sec4} the trivial pairs drop from the cohomology, i.e. the cohomological problem boils down to the cohomology of $s$ on functions $\omega(c,\xi,\tilde T)$ of the ghosts and the generalized tensors. This cohomology is nothing but $H(s_\mathrm{\,susy})$ because one has $s=s_\mathrm{\,susy}$ on these functions according to subsection \ref{sec6.4}.

To compute $H(s_\mathrm{\,susy})$ we apply the strategy outlined in section \ref{sec3} which starts from computing $H(s_\mathrm{gh})$ in the space of polynomials $f(c,\xi)$ of the ghosts. 
For the SUSY algebra \eqref{2Dalg} this cohomology is extremely simple and represented just by polynomials which are at most quadratic in the supersymmetric ghosts with the quadratic part proportional to the product $\xi^+\xi^-$ and which do not depend on the translation ghosts at all \cite{Brandt:2010fa}:
\begin{align}
&s_\mathrm{gh}\,f(c,\xi)=0\ \Leftrightarrow\ f(c,\xi)\sim a+\xi^+a_++\xi^-a_-+\xi^+\xi^-a_{+-}\,;\label{c1}\\
&a+\xi^+a_++\xi^-a_-+\xi^+\xi^-a_{+-}\sim 0\ \Rightarrow\ a=a_+=a_-=a_{+-}=0
\label{c2}
\end{align}
where $\sim$ denotes equivalence in $H(s_\mathrm{gh})$, i.e. $f\sim g\ \Leftrightarrow\ f=g+s_\mathrm{gh} h$.
This implies that the analysis of the SUSY ladder equations \eqref{ladder} in the present case is nontrivial only in $c$-degree\ zero \cite{Brandt:2009xv}. In particular one concludes that any nontrivial representative $\omega(c,\xi,\tilde T)$ of $H(s_\mathrm{\,susy})$ can be assumed to have a nonvanishing part $\omega^0$ with $c$-degree\ zero of the form
\begin{equation}
\omega^0=a(\tilde T)+\xi^+a_+(\tilde T)+\xi^-a_-(\tilde T)+\xi^+\xi^-a_{+-}(\tilde T).
\label{c3}
\end{equation}
The second equation in \eqref{ladder} (for $m=0$) then imposes that $d_\xi\omega^0$ must be $s_\mathrm{gh}$-exact, i.e.
\begin{equation}
d_\xi a(\tilde T)+\xi^+d_\xi a_+(\tilde T)+\xi^-d_\xi a_-(\tilde T)+\xi^+\xi^-d_\xi a_{+-}(\tilde T)=s_\mathrm{gh}(\dots).
\label{c4}
\end{equation}
Using now $d_\xi=\xi^+Q_++\xi^-Q_-$ one concludes by means of \eqref{c2} that the functions $a(\tilde T)$ and $a_\pm(\tilde T)$ must fulfill
\begin{equation}
Q_+a(\tilde T)=Q_-a(\tilde T)=Q_- a_+(\tilde T)+Q_+ a_-(\tilde T)=0.
\label{c5}
\end{equation}
These are the only obstructions imposed by the ladder equations on $\omega^0$ because, as mentioned already, the analysis of the SUSY ladder equations in the present case is nontrivial only in $c$-degree\ zero, i.e. any function $\omega^0$ of the form \eqref{c3} which fulfills \eqref{c5} gives rise to a cocycle of $H(s_\mathrm{\,susy})$. 
One can show that the obstructions \eqref{c5} imply that $a$ does not depend on generalized tensors at all. One now straightforwardly concludes the following result on $H(s_\mathrm{\,susy})$ in the present case stating that $H(s_\mathrm{\,susy})$ is trivial in all ghost numbers larger than 2 and is represented in ghost numbers 0, 1 and 2 by a constant and certain functions arising from the parts $\xi^+a_+(\tilde T)+\xi^-a_-(\tilde T)$ and $\xi^+\xi^-a_{+-}(\tilde T)$ of $\omega^0$, respectively:
\begin{align}
&s_\mathrm{\,susy}\omega(c,\xi,\tilde T)=0\ \Leftrightarrow\ \notag\\
&\omega=s_\mathrm{\,susy}\eta(c,\xi,\tilde T)+a
       + (\xi^++\mathrm{i}\,c^+Q_+)a_+(\tilde T)
       +(\xi^--\mathrm{i}\,c^-Q_-)a_-(\tilde T)\notag\\
&\phantom{\omega=}+(\xi^+\xi^-+\mathrm{i}\,c^+\xi^-Q_+-\mathrm{i}\,c^-\xi^+Q_--c^+c^-Q_+Q_-)a_{+-}(\tilde T) 
\label{c6}
\end{align}
where $a$ is a pure number and $a_\pm(\tilde T)$ fulfill \eqref{c5}, i.e. $Q_- a_+(\tilde T)+Q_+ a_-(\tilde T)=0$.

Furthermore the coboundary condition of $H(s_\mathrm{\,susy})$ provides at $c$-degree\ zero that $(\xi^++\mathrm{i}\,c^+Q_+)a_+(\tilde T)+(\xi^--\mathrm{i}\,c^-Q_-)a_-(\tilde T)$ is trivial in $H(s_\mathrm{\,susy})$ iff $a_+(T)=Q_+b(\tilde T)$ and $a_-(T)=Q_-b(\tilde T)$ with the same $b(\tilde T)$, and that  $(\xi^+\xi^-+\mathrm{i}\,c^+\xi^-Q_+-\mathrm{i}\,c^-\xi^+Q_--c^+c^-Q_+Q_-)a_{+-}(\tilde T)$ is trivial in $H(s_\mathrm{\,susy})$ iff
$a_{+-}(\tilde T)=Q_-b_+(\tilde T)+Q_+b_-(\tilde T)$ for some $b_\pm(\tilde T)$.

\subsection{Solutions to the descent equations}\label{sec6.6}

As we have discussed in section \ref{sec5}, the representatives of the supersymmetric BRST cohomology $H(s)$ directly give rise to representatives of $H(s+d)$ via the substitution $c\rightarrow c+dx$. The representatives of $H(s+d)$ provide solutions to the descent equations \eqref{descent} by decompsing them into parts with definite form-degrees. In order to illustrate this feature of the supersymmetric BRST cohomology we shall now briefly discuss it for the model under consideration.

As shown in subsection \ref{sec6.5}, $H(s)$ in this case only has nontrivial representatives with ghost numbers 0, 1 and 2. 

The representatives of $H(s)$ with ghost number 0 are just pure numbers, i.e. they do not provide solutions to the descent equations at form degrees different from zero. 

The representatives of $H(s)$ with ghost number 1 arise from functions $a_\pm(\tilde T)$ fulfilling $Q_- a_+(\tilde T)+Q_+ a_-(\tilde T)=0$ where $a_\pm(\tilde T)$ are not of the trivial form $a_+(T)=Q_+b(\tilde T)$, $a_-(T)=Q_-b(\tilde T)$ with the same $b(\tilde T)$. These representatives provide nontrivial solutions to the descent equations involving forms $\omega_p^{1-p}$ ($p=0,1,2$) where the subscript denotes the form-degree and the superscript denotes the ghost number. Recall that the BRST transformations under study only involve supersymmetry and translational transformations (but no other symmetries). According to general features of the local BRST cohomology \cite{Barnich:1994db,Brandt:1997cz}, the forms $\omega_2^{-1}$ and $\omega_1^{0}$ obtained from the representatives of $H(s)$ with ghost number 1 provide symmetries of the model under consideration which commute with the supersymmetry transformations $\delta_{\underline{\alpha}}$ and the translations $\delta_a$ at least on-shell, and the corresponding Noether currents, respectively. To give a simple example, we consider $a_\pm(T)=\pm\psi_{\pm(0,0)}$. It can be readily checked that these $a_\pm$ indeed fulfill $Q_- a_++Q_+ a_-=0$ and are not of the trivial form $a_+=Q_+b$ and $a_-=Q_-b$ with the same $b$. The representative $\omega$ of $H(s)$ with ghost number 1 arising from \eqref{c6} for this choice of $a_\pm$ and the corresponding forms $\omega_1^{0}$ and $\omega_2^{-1}$ read explicitly
\begin{align}
\omega&=\xi^+\psi_{+(0,0)}-\xi^-\psi_{-(0,0)}+\tfrac{1}{2}\,(c^+\varphi_{(1,0)}-c^-\varphi_{(0,1)})\, ,\label{s1}\\
\omega_1^0&=\tfrac{1}{2}\,(dx^+\partial_+-dx^-\partial_-)\varphi+\ldots=(dx^1\partial_2+dx^2\partial_1)\varphi+\ldots\, ,\label{s2}\\
\omega_2^{-1}&=-\tfrac{1}{2}\,dx^+dx^-\varphi^\star=dx^1dx^2\varphi^\star\label{s3}
\end{align}
with ellipsis indicating antifield dependent contributions to $\omega_1^0$.
\eqref{s3} provides the global symmetry of the action with Lagrangian \eqref{L} under constant shifts $\delta_{\mathrm{shift}}\varphi=1$ of the scalar field $\varphi$ (as the corresponding antifield $\varphi^\star$ in \eqref{s3} is multiplied just with the number 1 except for the volume element $dx^1dx^2$). This shift symmetry indeed commutes with the supersymmetry transformations $\delta_{\underline{\alpha}}$ and with the translations $\delta_a$. The antifield independent part of \eqref{s2} provides the corresponding Noether current $j^a=\partial^a\varphi$ written as the 1-form $dx^a\epsilon_{ab}j^b=dx^1 j_2+dx^2 j_1$ (with $\epsilon_{12}=1$).

The representatives of $H(s)$ with ghost number 2 arise from functions $a_{+-}(\tilde T)$ which are not of the trivial form $a_{+-}(\tilde T)=Q_-b_+(\tilde T)+Q_+b_-(\tilde T)$. These representatives provide nontrivial solutions to the descent equations involving forms $\omega_p^{2-p}$ ($p=0,1,2$). For the model under study the forms $\omega_2^{0}$ provide consistent first order deformations of its symmetries contained in $s$ (i.e., of the supersymmetry transformations $\delta_{\underline{\alpha}}$ and/or the translations $\delta_a$) and the corresponding first order deformations of the Lagrangian \eqref{L} \cite{Brandt:1997cz,Brandt:1998gj}. Again, we give a simple example for the purpose of illustration. As an example we choose $a_{+-}(\tilde T)=f(\varphi_{(0,0)})=f(\varphi)$ where $f(\varphi)$ is a function of the scalar field $\varphi$. As no function $f(\varphi)$ is of the form $Q_-b_+(\tilde T)+Q_+b_-(\tilde T)$ any choice $f(\varphi)$ provides a nontrivial representative of $H(s)$. The representative $\omega$ of $H(s)$ with ghost number 2 arising from \eqref{c6} for this choice of $a_{+-}$ and the corresponding form $\omega_2^{0}$ read explicitly
\begin{align}
\omega=&\,\xi^+\xi^-f(\varphi)+\mathrm{i}\,(c^+\xi^-\psi_{+(0,0)}-c^-\xi^+\psi_{-(0,0)})f^\prime(\varphi)\notag\\
&-c^+c^-\psi_{+(0,0)}\psi_{-(0,0)}f^{\prime\prime}(\varphi)\, ,\label{s4}\\
\omega_2^0=&\,dx^1dx^2[2\psi_+\psi_-f^{\prime\prime}(\varphi)+\tfrac{1}{2}\,(\psi^{\star -}\xi^+-\psi^{\star +}\xi^-)f^\prime(\varphi)]\label{s5}
\end{align}
where $f^\prime(\varphi)$ and $f^{\prime\prime}(\varphi)$ denote the first and second derivative of $f(\varphi)$ with respect to $\varphi$, respectively. The antifield independent part of \eqref{s5} provides a nontrivial first order deformation $L^{(1)}=2\psi_+\psi_-f^{\prime\prime}(\varphi)$ of the Lagrangian \eqref{L}. The antifield dependent part yields the corresponding first order deformations $\delta^{(1)}_{\underline{\alpha}}$ of the supersymmetry transformations with $\delta^{(1)}_\mp\psi_\pm=\pm\tfrac{1}{2} f^\prime(\varphi)$. Notice that the quadratic part of $f(\varphi)$ introduces a mass term for $\psi$ (the mass term for $\varphi$ arises in the corresponding second order deformation $L^{(2)}$ of the Lagrangian) and that terms in $f(\varphi)$ of higher order in $\varphi$ provide interactions of $\psi$ and $\varphi$.

Of course, there are many more nontrivial first order deformations of the Lagrangian \eqref{L} and its symmetries $\delta_{\underline{\alpha}}$ and $\delta_a$. For instance, one may replace or complement $f(\varphi)$ by a contribution $\varphi_{(1,0)}\psi_{+(0,0)}\varphi_{(0,1)}\psi_{-(0,0)}$ to $a_{+-}(\tilde T)$ which gives rise to an additional or complementary contribution to $L^{(1)}$ given by
$
2(\tfrac{\mathrm{i}}{2}\,\partial_+\varphi\partial_+\varphi+\psi_+\partial_+\psi_+)(\tfrac{\mathrm{i}}{2}\,\partial_-\varphi\partial_-\varphi-\psi_-\partial_-\psi_-)
$
and corresponding first oder deformations of the symmetry transformations.

\section{Concluding remarks}\label{sec7}

As we have discussed in section \ref{sec4} and demonstrated in section \ref{sec6}, a SAC $H(s_\mathrm{\,susy})$ as defined in section \ref{sec2} emerges within the supersymmetric BRST cohomology through an ``elimination of trivial pairs" from the cohomological analysis which eliminates derivatives of ghosts (if the ghosts are ghost fields) and antifields. Thereby the SAC emerges whether or not the commutator algebra of the symmetry transformations closes off-shell because the field variables which vanish on-shell are eliminated from the cohomological analysis along with the antifields. We remark that for the same reason it does not matter to the cohomological analysis whether or not one uses auxiliary fields (if any) to close the algebra of the symmetry transformations because these auxiliary fields are also eliminated from the cohomological analysis along with their antifields in this approach. 

As we have also discussed and demonstrated in some detail, the elimination of the antifields leads to a cohomological problem involving only ghost variables and ``generalized tensors" $\tilde T=T+\dots$ where the dots indicate possible antifield dependent contributions.  The respective SUSY algebra \eqref{alg} is represented on these generalized tensors with an unusual representation of the translational generators $P_a$ corresponding to a representation of derivatives on-shell.

Of course one may use an alternative approach and keep the antifields throughout the cohomological analysis which leads to a cohomological problem involving ghosts, standard tensors $T$ and antitensors $T^\star$ corresponding to the antifields and their derivatives \cite{Barnich:1995ap}. In that approach the cohomological analysis gives rise to a variant of the SAC which in general involves an on-shell representation of the respective SUSY algebra \eqref{alg} \cite{Brandt:1996au}.

The SAC related to a SUSY algebra \eqref{alg} is the counterpart of a Lie algebra cohomology $H(s_\mathrm{\,Lie})$ related to a Lie algebra $[\, \delta_i\, ,\, \delta_j\, ]=f_{ij}{}^k\delta_k$ with a BRST-type differential $s_\mathrm{\,Lie}=C^i\delta_i+\tfrac{1}{2} C^jC^kf_{kj}{}^i\frac{\partial}{\partial C^i}$ acting on ghosts $C$ and tensors $T$. Now, in spite of the similiar form of $s_\mathrm{\,susy}$ and $s_\mathrm{\,Lie}$, the structures of $H(s_\mathrm{\,susy})$ and $H(s_\mathrm{\,Lie})$ differ considerably. Namely, $H(s_\mathrm{\,Lie})$ (for semisimple Lie algebras) factorizes with respect to the ghosts and tensors in the sense that its representatives take the form $\sum_r f^r(C)g_r(T)$ with $s_\mathrm{\,Lie}\, f^r(C)=0$ and $s_\mathrm{\,Lie}\, g_r(T)=0$. 
In sharp contrast, $H(s_\mathrm{\,susy})$ does not factorize in this way because usually there are no nontrivial $s_\mathrm{\,susy}$-invariant functions $g(\tilde T)$ owing to the presence of the translational generators $P_A$ in \eqref{alg}. 

However, the representatives $f(C)$ of $H(s_\mathrm{\,Lie})$ do have counterparts in $H(s_\mathrm{\,susy})$ which are polynomials $f(c,\xi)$ in the translation ghosts $c$ and the supersymmetry ghosts $\xi$ representing the cohomology $H(s_\mathrm{gh})$ of the part $s_\mathrm{gh}$ of $s_\mathrm{\,susy}$ given in \eqref{gen8}. This part acts nontrivially only on the ghosts and is the counterpart of the part $\tfrac{1}{2} C^jC^kf_{kj}{}^i\frac{\partial}{\partial C^i}$ of $s_\mathrm{\,Lie}$. It 
plays a distinguished role within SAC because it is the only part of $s_\mathrm{\,susy}$ which decrements the degree in the translation ghosts. In particular this allows one to relate $H(s_\mathrm{\,susy})$ to $H(s_\mathrm{gh})$ by means of spectral sequence techniques \cite{Brandt:2009xv}, as we have sketched in section \ref{sec3}. 
$H(s_\mathrm{gh})$ has been computed recently for various SUSY algebras \eqref{alg} in diverse dimensions \cite{Brandt:2010fa,Brandt:2010tz,Movshev:2010mf,Movshev:2011pr}.
\\

{\it Acknowledgment:} I would like to thank the organizers of the conference in memory of Maximilian Kreuzer for inviting me to the conference and for giving me the opportunity to
contribute to the Memorial Volume for him.

\end{document}